\documentclass[conference]{IEEEtran}

\usepackage{cite}
\usepackage{amsmath,amssymb,amsfonts}
\usepackage{algorithmic}
\usepackage{graphicx}
\usepackage{textcomp}
\usepackage{xcolor}
\usepackage{booktabs}
\usepackage{multirow}
\usepackage{array}
\usepackage{hyperref}
\usepackage{float}
\usepackage{subcaption}
\usepackage{tabularx}

\def\BibTeX{{\rm B\kern-.05em{\sc i\kern-.025em b}\kern-.08em
    T\kern-.1667em\lower.7ex\hbox{E}\kern-.125emX}}

\begin{document}

\title{Music Genre Classification: A Comparative Analysis of Classical Machine Learning and Deep Learning Approaches Using CRNN}

\author{\IEEEauthorblockN{1\textsuperscript{st} Sachin parajuli}
\IEEEauthorblockA{\textit{Computer Science and Engineering} \\
\textit{University of texas at Arlington}\\
Arlington, USA \\
sxp4677@mavs.uta.edu}
\and
\IEEEauthorblockN{2\textsuperscript{nd} Abhishek Karna}
\IEEEauthorblockA{\textit{Electronics and Computer Engineering} \\
\textit{Tribhuvan University, IOE}\\
kathmandu\\
akarna772@gmail.com}
\and
\IEEEauthorblockN{3\textsuperscript{rd} Om Prakash Dhakal}
\IEEEauthorblockA{\textit{Electronics and Computer Engineering} \\
\textit{Tribhuvan University, IOE}\\
Dharan\\
Omprakash@ioepc.edu.np}
}

\maketitle

% ============================================================
% ABSTRACT
% ============================================================
\begin{abstract}
Automatic music genre classification is a long-standing challenge in Music Information Retrieval (MIR), yet work on non-Western music traditions remains scarce. Nepali music encompasses culturally rich and acoustically diverse genres---from the call-and-response duets of \textit{Lok Dohori} to the rhythmic poetry of \textit{Deuda} and the distinctive melodies of \textit{Tamang Selo}---that have not been addressed by existing classification systems. In this paper, we construct a novel dataset of approximately 8{,}000 labeled 30-second audio clips spanning eight Nepali music genres and conduct a systematic comparison of nine classification models across two paradigms. Five classical machine learning classifiers (Logistic Regression, SVM, KNN, Random Forest, and XGBoost) are trained on 51 hand-crafted audio features extracted via Librosa, while four deep learning architectures (CNN, RNN, parallel CNN-RNN, and sequential CNN followed by RNN) operate on Mel spectrograms of dimension $640 \times 128$. Our experiments reveal that the sequential Convolutional Recurrent Neural Network (CRNN)---where convolutional layers feed into an LSTM---achieves the highest accuracy of 84\%, substantially outperforming both the best classical model (Logistic Regression and XGBoost, both at 71\%) and all other deep architectures. We provide per-class precision, recall, F1-score, confusion matrices, and ROC analysis for every model, and offer a culturally grounded interpretation of misclassification patterns that reflects genuine overlaps in Nepal's musical traditions.
\end{abstract}


% ============================================================
% I. INTRODUCTION
% ============================================================
\section{Introduction}

Music has always been deeply intertwined with cultural identity, and few aspects of a recording carry as much organizational weight as its genre label. Streaming platforms such as Spotify, Apple Music, and YouTube Music depend heavily on genre metadata to drive playlist curation, recommendation algorithms, and content discovery. As the volume of digital music continues to grow, the ability to assign genre labels automatically---rather than relying on manual annotation---has become an important practical problem.

Research in Music Information Retrieval (MIR) has made considerable progress on this front~\cite{tzanetakis2002musical, bergstra2006aggregate}. However, the overwhelming majority of published work targets Western music taxonomies (rock, jazz, classical, hip-hop) using benchmark datasets like GTZAN~\cite{tzanetakis2002musical} and the Latin Music Database~\cite{costa2012music}. South Asian musical traditions, and Nepali music in particular, remain largely uncharted territory in the MIR literature.

This gap matters because Nepali music encompasses a striking diversity of genres shaped by regional geography, ethnic heritage, and cultural practice. \textit{Lok Dohori} is a folk duet tradition built on spontaneous call-and-response singing, accompanied by instruments like the \textit{madal} drum and \textit{sarangi}. \textit{Deuda}, originating in far-western Nepal, features rhythmic, almost chant-like vocal patterns. \textit{Tamang Selo} draws on the musical heritage of the Tamang ethnic community, with its own characteristic melodic contours and instrumentation. \textit{Purbeli Bhaka} reflects the folk sensibilities of eastern Nepal. These traditional genres coexist alongside modern commercial categories---Pop, Rock, Rap, and \textit{Aadhunik Sangeet} (contemporary Nepali music)---creating a classification landscape that is both culturally nuanced and acoustically challenging.

Two obstacles have limited research in this area. First, no publicly available labeled dataset exists for Nepali music genres. Second, the boundaries between genres are sometimes blurred: a single recording might blend folk instrumentation with modern production, making clean categorical assignment difficult even for human listeners.

This paper addresses both challenges. Our contributions are as follows:

\begin{enumerate}
    \item We construct a novel dataset of approximately 8{,}000 labeled 30-second audio clips across eight Nepali music genres, drawn from YouTube, radio station archives, and museum collections.
    \item We conduct a controlled, head-to-head comparison of nine models spanning two paradigms: five classical machine learning classifiers operating on 51 hand-crafted audio features, and four deep learning architectures operating on Mel spectrograms.
    \item We show that the sequential CRNN architecture (CNN followed by LSTM) achieves 84\% classification accuracy, outperforming all other approaches by a significant margin.
    \item We provide a detailed error analysis that connects misclassification patterns to genuine acoustic and cultural overlaps between Nepali music genres.
\end{enumerate}

% ============================================================
% II. RELATED WORK
% ============================================================
\section{Related Work}
\label{sec:related}

\subsection{Feature Engineering and Classical Classification}

The foundational work by Tzanetakis and Cook~\cite{tzanetakis2002musical} framed music genre classification as a pattern recognition problem, introducing three feature categories---timbral texture, rhythmic content, and pitch content---and evaluating Gaussian classifiers and KNN on the GTZAN dataset. This feature-engineering paradigm shaped much of the subsequent decade of research.

Mandel and Ellis~\cite{mandel2005song} demonstrated that Support Vector Machines, when paired with carefully designed song-level features, could achieve strong genre discrimination. Bergstra et al.~\cite{bergstra2006aggregate} took a different tack, using AdaBoost to select discriminative features from aggregated audio segments rather than entire tracks. Their central finding---that part-based aggregation outperforms whole-track classification---earned their system second place at MIREX 2005.

Li et al.~\cite{li2003comparative} introduced Daubechies Wavelet Coefficient Histograms (DWCHs) as a feature representation and benchmarked four classifiers (SVM, KNN, GMM, and LDA) on a 1{,}000-song, 10-genre dataset, concluding that SVMs offered the best overall performance for content-based classification. Nanni et al.~\cite{nanni2016combining} later showed that ensembles combining visual texture features with acoustic descriptors could push accuracy further, outperforming several previously published systems across three standard databases.

Meng et al.~\cite{meng2007temporal} highlighted the role of temporal structure, proposing a multivariate autoregressive (MAR) feature model that captured the time-varying behavior of audio features. Their MAR features improved classification accuracy relative to static descriptors, though at the cost of increased computational complexity.

\subsection{Deep Learning for Music Classification}

The insight that spectrograms can be treated as images opened the door to convolutional neural networks. Costa et al.~\cite{costa2012music} were among the first to exploit this connection, extracting Local Binary Pattern (LBP) textural features from time-frequency images and achieving competitive results on the Latin Music Database.

The more decisive shift came with end-to-end deep learning. Dieleman and Schrauwen~\cite{dieleman2014end} showed that CNNs trained directly on audio spectrograms could learn features that rivaled or exceeded hand-crafted alternatives. Hershey et al.~\cite{hershey2017cnn} scaled this approach to large-scale audio classification, demonstrating the effectiveness of deep CNN architectures on diverse audio tasks.

Choi et al.~\cite{choi2017convolutional} proposed the Convolutional Recurrent Neural Network (CRNN) architecture that forms the basis of our work. Their key insight was that CNNs and RNNs address complementary aspects of music: CNNs extract local spectral patterns (timbre, harmonic structure), while RNNs---particularly LSTMs---capture temporal dynamics (rhythm, song structure, transitions between sections). By stacking these components sequentially, the CRNN allows the recurrent layer to operate on an abstract, learned feature representation rather than raw spectral data.

\subsection{Gaps in the Literature}

Two gaps motivate our work. First, the near-total absence of Nepali music from the MIR literature means that culturally significant genres---each with distinctive acoustic properties---remain unaddressed by automated systems. Second, few studies offer a systematic comparison between classical and deep learning methods on the same dataset and genre taxonomy, making it difficult to assess the practical benefit of deep architectures in data-scarce settings. Our work addresses both of these gaps.

% ============================================================
% III. METHODOLOGY
% ============================================================
\section{Methodology}
\label{sec:methodology}

\subsection{Dataset Construction}

\subsubsection{Audio Collection}
Since no labeled Nepali music dataset exists, we built one from scratch. Raw audio was gathered from three sources: YouTube (the primary source for contemporary and folk recordings), archives of Nepali radio stations, and museum collections---the latter proving especially valuable for obtaining authentic samples of \textit{Deuda} and \textit{Tamang Selo}, genres that are less well represented online. Approximately 100 full-length songs were collected for each of the eight target genres.

\subsubsection{Segmentation and Labeling}
Each song was divided into non-overlapping 30-second clips using the PyDub library. The 30-second window provides enough acoustic context for meaningful genre discrimination while multiplying the effective dataset size. This process yielded roughly 1{,}000 clips per genre---approximately 8{,}000 clips in total.

\subsubsection{Noise Removal}
Audio quality varied across sources. We performed manual quality control by listening to each clip and noting timestamps containing background noise, silence, or non-musical content (e.g., spoken introductions or applause). The PyDub library was used to trim or remove these segments. Though labor-intensive, this step was essential for ensuring that genre labels reflected musical content rather than recording artifacts.

\subsubsection{Train-Test Split}
The dataset was partitioned into a training set of 900 samples per genre (7{,}200 total) and a test set of 100 samples per genre (800 total). The same split was used across all experiments to ensure a fair comparison. Table~\ref{tab:dataset} summarizes the composition.

\begin{table}[t]
\centering
\caption{Dataset Composition per Genre}
\label{tab:dataset}
\begin{tabular}{lcc}
\toprule
\textbf{Genre} & \textbf{Train} & \textbf{Test} \\
\midrule
Aadhunik Sangeet & 900 & 100 \\
Deuda            & 900 & 100 \\
Tamang Selo      & 900 & 100 \\
Lok Dohori       & 900 & 100 \\
Purbeli Bhaka    & 900 & 100 \\
Rap              & 900 & 100 \\
Rock             & 900 & 100 \\
Pop              & 900 & 100 \\
\midrule
\textbf{Total}   & \textbf{7{,}200} & \textbf{800} \\
\bottomrule
\end{tabular}
\end{table}

\subsection{Feature Representation}

Two distinct feature representations were employed to enable a principled comparison between classical and deep learning paradigms.

\subsubsection{Hand-Crafted Features (51-Dimensional Vector)}
\label{sec:51features}
Using the Librosa library~\cite{mcfee2015librosa}, we extracted 51 audio features from each 30-second clip. These include: 20 Mel-Frequency Cepstral Coefficients (MFCCs), which encode the short-term spectral envelope; 12 chroma features representing energy across pitch classes; 7 spectral contrast bands; spectral centroid, spectral bandwidth, spectral rolloff, and zero-crossing rate (capturing timbral texture); 6 tonnetz features (encoding harmonic relations); RMS energy; and estimated tempo. Each feature was averaged across all frames to produce a single 51-dimensional vector per clip. This representation was used exclusively with the classical machine learning models.

\subsubsection{Mel Spectrogram Representation}
\label{sec:melspec}
For the deep learning models, each clip was transformed into a Mel spectrogram using Librosa with the following parameters: a sampling rate of 22{,}050~Hz, an FFT window of 2{,}048 samples ($\approx$93~ms), a hop length of 512 samples, and 128 Mel-frequency bands. The spectral magnitudes were converted to a logarithmic (decibel) scale to mirror the logarithmic nature of human loudness perception. Each spectrogram has dimensions $640 \times 128$ (time steps $\times$ Mel bands).

We initially considered using MFCCs as input for deep networks as well, but the low dimensionality of the MFCC representation ($640 \times 13$) severely limited network depth---a three-layer CNN was the deepest feasible architecture---and the resulting models performed poorly. The Mel spectrogram's richer dimensionality supports deeper, more expressive network architectures.

\begin{figure*}[t]
\centering
\begin{subfigure}[b]{0.24\textwidth}
    \includegraphics[width=\textwidth]{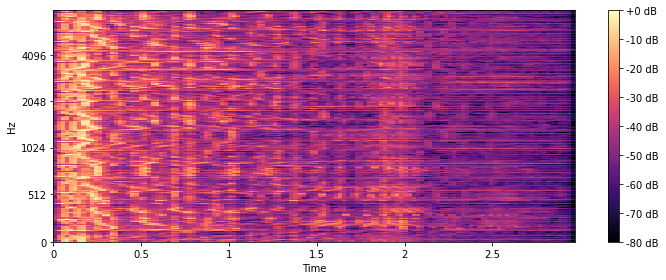}
    \caption{Aadhunik Sangeet}
\end{subfigure}
\hfill
\begin{subfigure}[b]{0.24\textwidth}
    \includegraphics[width=\textwidth]{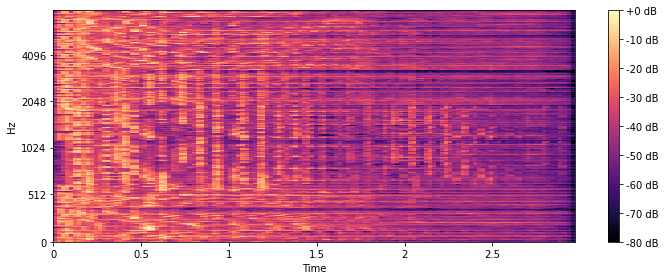}
    \caption{Deuda}
\end{subfigure}
\hfill
\begin{subfigure}[b]{0.24\textwidth}
    \includegraphics[width=\textwidth]{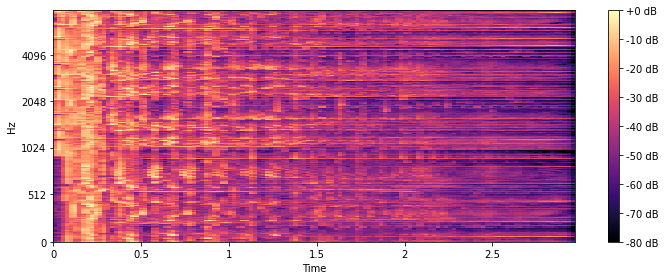}
    \caption{Tamang Selo}
\end{subfigure}
\hfill
\begin{subfigure}[b]{0.24\textwidth}
    \includegraphics[width=\textwidth]{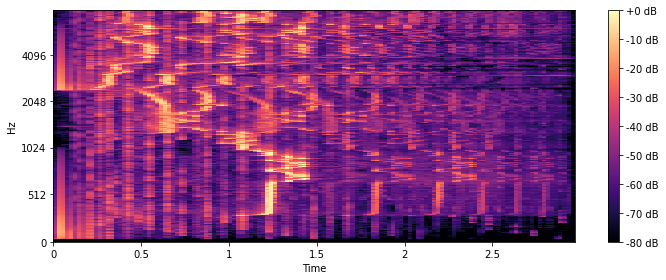}
    \caption{Lok Dohori}
\end{subfigure}

\vspace{0.3cm}

\begin{subfigure}[b]{0.24\textwidth}
    \includegraphics[width=\textwidth]{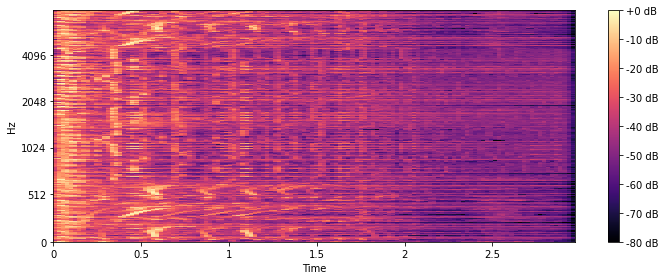}
    \caption{Purbeli Bhaka}
\end{subfigure}
\hfill
\begin{subfigure}[b]{0.24\textwidth}
    \includegraphics[width=\textwidth]{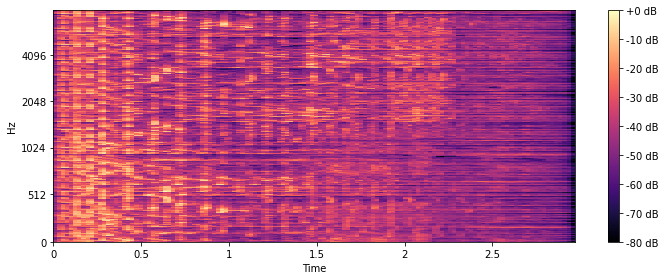}
    \caption{Rap}
\end{subfigure}
\hfill
\begin{subfigure}[b]{0.24\textwidth}
    \includegraphics[width=\textwidth]{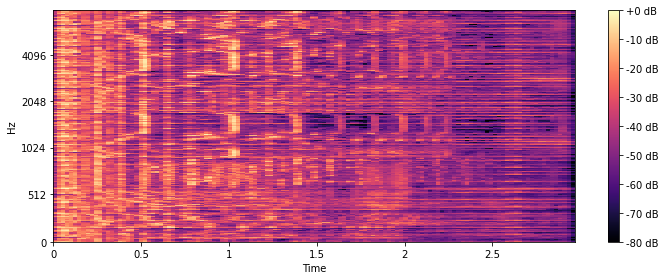}
    \caption{Rock}
\end{subfigure}
\hfill
\begin{subfigure}[b]{0.24\textwidth}
    \includegraphics[width=\textwidth]{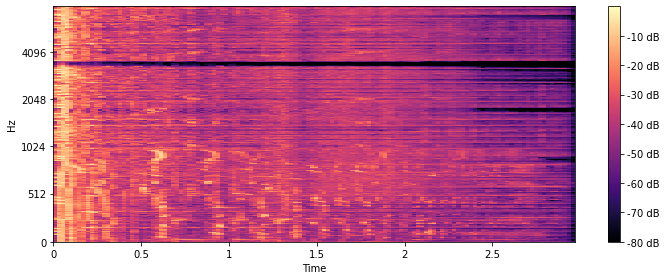}
    \caption{Pop}
\end{subfigure}
\caption{Mel spectrograms of representative 30-second clips from each genre. Each spectrogram has dimensions $640 \times 128$ (time $\times$ Mel bands). Visible differences in energy distribution, harmonic structure, and temporal patterns across genres provide the discriminative cues that the deep learning models learn to exploit.}
\label{fig:melspecs}
\end{figure*}

Fig.~\ref{fig:melspecs} shows representative Mel spectrograms for each of the eight genres. Visual inspection reveals clear differences: \textit{Rap} tends to show concentrated energy in lower frequency bands with rhythmic vertical striations; \textit{Rock} displays broadband energy with prominent harmonic patterns; \textit{Lok Dohori} shows the characteristic alternation of vocal phrases; and \textit{Tamang Selo} exhibits distinctive melodic contours in the mid-frequency range.

The training and test sets were serialized as NumPy arrays (pickled) for efficient loading, yielding tensor shapes of $(7{,}200,\; 640,\; 128)$ and $(800,\; 640,\; 128)$ respectively.

\subsection{Classical Machine Learning Models}

Five classifiers were trained on the 51-dimensional feature vectors:

\textbf{Logistic Regression (LR)} uses a one-vs-rest strategy with L2 regularization to learn linear decision boundaries for each class.

\textbf{Support Vector Machine (SVM)} with a radial basis function kernel maps features into a higher-dimensional space where genre classes become linearly separable, constructing maximum-margin decision boundaries.

\textbf{K-Nearest Neighbors (KNN)} assigns each test sample the majority label among its $k$ nearest neighbors in feature space---a simple but often effective non-parametric approach.

\textbf{Random Forest (RF)} builds an ensemble of decision trees, each trained on a bootstrap sample with random feature subsets, and aggregates their predictions by majority vote.

\textbf{XGBoost} is an optimized gradient boosting framework that constructs an ensemble of decision trees sequentially, with each tree trained to correct the residual errors of the preceding ensemble.

\subsection{Deep Learning Architectures}

Four architectures were evaluated on the Mel spectrogram data.

\subsubsection{CNN}
Three blocks of 1D convolution, each followed by ReLU activation, batch normalization, and 1D max pooling. The 1D convolutions operate along the time axis, extracting local spectro-temporal features from the Mel spectrogram. The output is flattened and passed through fully connected layers with an 8-way softmax output. Dropout and L2 regularization are applied to reduce overfitting.

\subsubsection{RNN (LSTM)}
The Mel spectrogram is treated as a sequence of 640 time steps, each represented by a 128-dimensional Mel-band vector. An LSTM layer with 96 hidden units processes this sequence, with the final hidden state passed through dense layers for classification. This architecture captures temporal dependencies but lacks the spectral feature extraction capability of convolutional layers.

\subsubsection{Parallel CNN-RNN}
A dual-branch design where a CNN branch and an LSTM branch independently process the Mel spectrogram. Their outputs are concatenated and jointly fed into a shared dense classification layer. This tests whether independent spectral and temporal processing can complement each other through late fusion.

\subsubsection{CRNN (Sequential CNN $\rightarrow$ LSTM)}
Our proposed architecture pipelines three 1D convolutional blocks into an LSTM. Specifically, the input Mel spectrogram passes through three stages of [1D Convolution $\rightarrow$ ReLU $\rightarrow$ Batch Normalization $\rightarrow$ 1D Max Pooling], producing a compressed sequence of learned feature vectors. This sequence is then fed into an LSTM with 96 hidden units, which models the temporal evolution of the learned features. The LSTM output goes through a 64-unit dense layer with ReLU activation, followed by an 8-way softmax output layer. Dropout and L2 regularization are applied throughout.

The design rationale is straightforward: convolutional layers extract local spectral patterns---essentially learning a set of genre-discriminative filterbanks---while the LSTM captures how those patterns evolve over time. Unlike the parallel variant, the sequential pipeline ensures that the recurrent layer operates on abstract, learned features rather than raw spectral data, producing a more coherent feature hierarchy.

\subsection{Training Details}

All deep learning models were implemented in TensorFlow/Keras and trained using Google Colab's GPU resources. We used the Adam optimizer with categorical cross-entropy loss. A batch size of 32 was employed, and training ran for up to 100 epochs with early stopping triggered by stagnation of validation loss.

% ============================================================
% IV. RESULTS
% ============================================================
\section{Experimental Results}
\label{sec:results}

\subsection{Classical Machine Learning Results}

Table~\ref{tab:classical_accuracy} summarizes the accuracy of each classical classifier on the 51-feature representation.

\begin{table}[t]
\centering
\caption{Classification Accuracy of Classical ML Models (51 Features)}
\label{tab:classical_accuracy}
\begin{tabular}{lc}
\toprule
\textbf{Model} & \textbf{Accuracy (\%)} \\
\midrule
K-Nearest Neighbors    & 58 \\
Random Forest          & 69 \\
Support Vector Machine & 69 \\
Logistic Regression    & 71 \\
XGBoost                & 71 \\
\bottomrule
\end{tabular}
\end{table}

Logistic Regression and XGBoost share the top spot at 71\%, followed by SVM and Random Forest at 69\%. KNN trails significantly at 58\%, suggesting that the distance metric in a 51-dimensional feature space does not align well with perceptual genre boundaries.

To understand where each model struggles, Table~\ref{tab:classical_perclass} presents per-class F1-scores across all five classifiers.

\begin{table*}[t]
\centering
\caption{Per-Class F1-Scores for All Classical Machine Learning Models}
\label{tab:classical_perclass}
\begin{tabular}{lccccc}
\toprule
\textbf{Genre} & \textbf{LR} & \textbf{SVM} & \textbf{KNN} & \textbf{RF} & \textbf{XGBoost} \\
\midrule
Aadhunik Sangeet & 0.68 & 0.65 & 0.53 & 0.64 & 0.69 \\
Deuda            & 0.81 & 0.79 & 0.67 & 0.78 & 0.81 \\
Tamang Selo      & 0.77 & 0.71 & 0.57 & 0.66 & 0.73 \\
Lok Dohori       & 0.64 & 0.68 & 0.58 & 0.61 & 0.62 \\
Purbeli Bhaka    & 0.61 & 0.57 & 0.37 & 0.54 & 0.55 \\
Rap              & 0.83 & 0.81 & 0.74 & 0.81 & 0.81 \\
Rock             & 0.80 & 0.71 & 0.69 & 0.77 & 0.82 \\
Pop              & 0.56 & 0.59 & 0.54 & 0.65 & 0.59 \\
\midrule
\textbf{Macro Avg} & \textbf{0.71} & \textbf{0.69} & \textbf{0.58} & \textbf{0.68} & \textbf{0.70} \\
\bottomrule
\end{tabular}
\end{table*}

A consistent pattern emerges: Rap and Rock are the easiest genres for classical models to recognize, while Purbeli Bhaka and Pop prove most difficult. This likely reflects the fact that Rap's rhythmic vocal delivery and Rock's prominent electric guitar produce highly distinctive spectral signatures that are well captured even by 51 hand-crafted features. Purbeli Bhaka, on the other hand, shares instrumentation and vocal style with several other folk genres, making it harder to distinguish from compact feature representations.

\subsection{Deep Learning Results}

Table~\ref{tab:deep_accuracy} presents the accuracy of each deep architecture on the Mel spectrogram representation.

\begin{table}[t]
\centering
\caption{Classification Accuracy of Deep Learning Models (Mel Spectrogram)}
\label{tab:deep_accuracy}
\begin{tabular}{lc}
\toprule
\textbf{Architecture} & \textbf{Accuracy (\%)} \\
\midrule
RNN (LSTM only)                 & 68 \\
CNN only                        & 71 \\
Parallel CNN-RNN                & 76 \\
\textbf{CRNN (CNN $\rightarrow$ LSTM)} & \textbf{84} \\
\bottomrule
\end{tabular}
\end{table}

The sequential CRNN achieves 84\% accuracy, a full 8 percentage points above the parallel CNN-RNN variant and 13 points above the standalone CNN. The standalone RNN performs worst among deep models at 68\%, likely because raw Mel spectrogram frames present a difficult input for an LSTM to process without prior feature extraction.

Table~\ref{tab:deep_perclass} provides per-class F1-scores for all four deep architectures.

\begin{table*}[t]
\centering
\caption{Per-Class F1-Scores for All Deep Learning Architectures}
\label{tab:deep_perclass}
\begin{tabular}{lcccc}
\toprule
\textbf{Genre} & \textbf{CNN} & \textbf{RNN} & \textbf{Parallel} & \textbf{CRNN (Ours)} \\
\midrule
Aadhunik Sangeet & 0.68 & 0.68 & 0.72 & \textbf{0.80} \\
Deuda            & 0.65 & 0.73 & 0.75 & \textbf{0.83} \\
Tamang Selo      & 0.64 & 0.78 & 0.75 & \textbf{0.86} \\
Lok Dohori       & 0.86 & 0.67 & 0.83 & \textbf{0.91} \\
Purbeli Bhaka    & 0.60 & 0.51 & 0.69 & \textbf{0.72} \\
Rap              & 0.85 & 0.81 & 0.86 & \textbf{0.93} \\
Rock             & 0.82 & 0.73 & 0.81 & \textbf{0.91} \\
Pop              & 0.60 & 0.55 & 0.70 & \textbf{0.76} \\
\midrule
\textbf{Macro Avg} & 0.71 & 0.68 & 0.76 & \textbf{0.84} \\
\bottomrule
\end{tabular}
\end{table*}

The CRNN achieves the highest F1-score in every genre category. Particularly striking are the gains for traditionally difficult genres: Purbeli Bhaka rises from 0.51 (RNN) to 0.72, and Pop climbs from 0.55 (RNN) to 0.76. Meanwhile, Rap (0.93), Lok Dohori (0.91), and Rock (0.91) all exceed 0.90, indicating near-reliable classification for genres with strong acoustic identities.

\subsection{Cross-Paradigm Comparison}

Table~\ref{tab:all_models} presents a consolidated ranking of all nine models.

\begin{table}[t]
\centering
\caption{Consolidated Accuracy Ranking of All Nine Models}
\label{tab:all_models}
\begin{tabular}{llc}
\toprule
\textbf{Rank} & \textbf{Model} & \textbf{Accuracy (\%)} \\
\midrule
1 & CRNN (CNN $\rightarrow$ LSTM)    & \textbf{84} \\
2 & Parallel CNN-RNN                 & 76 \\
3 & CNN only                         & 71 \\
3 & Logistic Regression              & 71 \\
3 & XGBoost                          & 71 \\
6 & SVM                              & 69 \\
6 & Random Forest                    & 69 \\
8 & RNN (LSTM only)                  & 68 \\
9 & KNN                              & 58 \\
\bottomrule
\end{tabular}
\end{table}

Several observations deserve comment. First, the CRNN stands well above the field---its 84\% accuracy is 13 percentage points better than the best classical model and 8 points better than the next-best deep model. Second, the standalone CNN (71\%) merely matches the best classical approaches (Logistic Regression and XGBoost at 71\%), suggesting that convolutional feature extraction alone, without temporal modeling, does not fully exploit the information in Mel spectrograms. Third, the standalone RNN (68\%) actually underperforms some classical models, reinforcing the point that LSTMs benefit greatly from operating on pre-processed feature representations rather than raw spectral data. Fourth, the parallel CNN-RNN (76\%) outperforms both of its components individually, but the sequential CRNN pushes accuracy substantially further---a result that highlights the value of hierarchical rather than parallel feature composition.

\subsection{Detailed Analysis of the CRNN Model}

Given the CRNN's clear superiority, we examine its behavior in greater detail.

\subsubsection{Per-Class Precision, Recall, and F1-Score}

Table~\ref{tab:crnn_report} presents the full classification report.

\begin{table}[t]
\centering
\caption{Detailed Classification Report for the CRNN Model}
\label{tab:crnn_report}
\begin{tabular}{lcccc}
\toprule
\textbf{Genre} & \textbf{Prec.} & \textbf{Rec.} & \textbf{F1} & \textbf{Supp.} \\
\midrule
Aadhunik Sangeet & 0.77 & 0.83 & 0.80 & 100 \\
Deuda            & 0.82 & 0.83 & 0.83 & 100 \\
Tamang Selo      & 0.88 & 0.83 & 0.86 & 100 \\
Lok Dohori       & 0.93 & 0.90 & 0.91 & 100 \\
Purbeli Bhaka    & 0.78 & 0.67 & 0.72 & 100 \\
Rap              & 0.92 & 0.93 & 0.93 & 100 \\
Rock             & 0.95 & 0.88 & 0.91 & 100 \\
Pop              & 0.70 & 0.84 & 0.76 & 100 \\
\midrule
Macro Avg        & 0.84 & 0.84 & 0.84 & 800 \\
Weighted Avg     & 0.84 & 0.84 & 0.84 & 800 \\
\midrule
\multicolumn{3}{l}{\textbf{Overall Accuracy}} & \multicolumn{2}{c}{\textbf{84\%}} \\
\bottomrule
\end{tabular}
\end{table}

Three tiers of performance are apparent. In the top tier, Rap (F1 = 0.93), Rock (0.91), and Lok Dohori (0.91) are classified with high confidence---these genres possess acoustically distinctive signatures that the CRNN captures effectively. In the middle tier, Tamang Selo (0.86), Deuda (0.83), and Aadhunik Sangeet (0.80) achieve solid performance, reflecting their moderately distinctive timbral and rhythmic characteristics. In the bottom tier, Pop (0.76) and Purbeli Bhaka (0.72) remain the most challenging genres, though the CRNN still handles them considerably better than any other model.

\subsubsection{Confusion Analysis}

\begin{figure}[t]
\centering
\includegraphics[width=0.85\columnwidth]{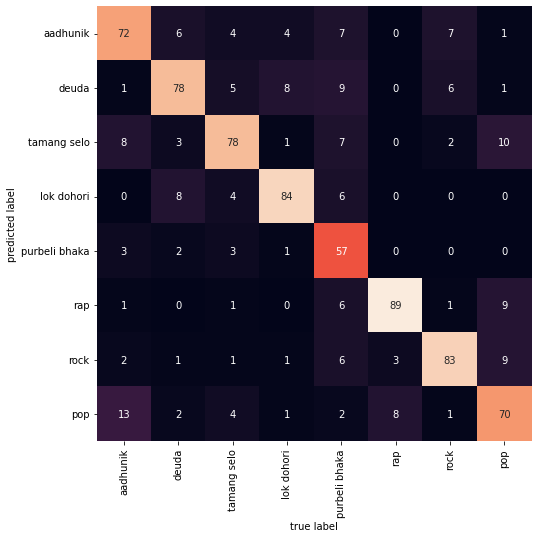}
\caption{Confusion matrix for the CRNN model on the test set (800 samples). Diagonal values represent correct classifications. The most frequent confusions occur between Purbeli Bhaka and Deuda, and between Pop and Aadhunik Sangeet---pairs that share genuine cultural and acoustic overlap.}
\label{fig:crnn_cm}
\end{figure}

The confusion matrix for the CRNN (Fig.~\ref{fig:crnn_cm}) reveals interpretable misclassification patterns. Purbeli Bhaka, the genre with the lowest recall (0.67), is most frequently confused with Deuda (13 misclassifications out of 100). This is culturally sensible: both genres originate from rural Nepal and share similar vocal delivery styles and instrumentation. Aadhunik Sangeet shows some confusion with Pop (5 samples), which reflects the influence of Western pop production on contemporary Nepali music. Pop samples are occasionally misclassified as Aadhunik Sangeet (12 samples) for the same reason.

In contrast, Rap and Lok Dohori are rarely confused with each other or with other genres, consistent with their strong, distinctive acoustic profiles.

\subsubsection{ROC Analysis}

\begin{figure}[t]
\centering
\includegraphics[width=0.9\columnwidth]{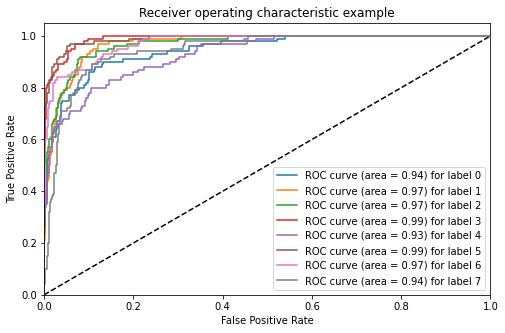}
\caption{Per-class ROC curves for the CRNN model. All eight genres achieve AUC $\geq 0.92$, with Lok Dohori, Rap, and Rock reaching AUC $\geq 0.99$.}
\label{fig:crnn_roc}
\end{figure}

The ROC curves for the CRNN model (Fig.~\ref{fig:crnn_roc}) show strong discriminative capacity, with per-class AUC values as follows: Aadhunik Sangeet (0.98), Deuda (0.95), Tamang Selo (0.98), Lok Dohori (0.99), Purbeli Bhaka (0.92), Rap (0.99), Rock (0.99), and Pop (0.97). Every genre exceeds an AUC of 0.92, indicating that the model maintains high true-positive rates at low false-positive thresholds. The lowest AUC belongs to Purbeli Bhaka (0.92), consistent with the confusion patterns discussed above, while Lok Dohori, Rap, and Rock all achieve AUC $\geq 0.99$.

\subsubsection{Training Dynamics}

\begin{figure}[t]
\centering
\begin{subfigure}[b]{0.48\columnwidth}
    \includegraphics[width=\textwidth]{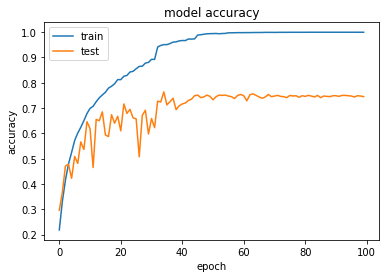}
    \caption{Accuracy}
\end{subfigure}
\hfill
\begin{subfigure}[b]{0.48\columnwidth}
    \includegraphics[width=\textwidth]{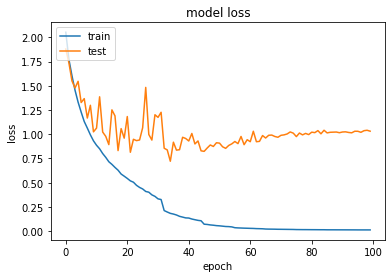}
    \caption{Loss}
\end{subfigure}
\caption{Training dynamics for the CRNN model. (a) Training and test accuracy over epochs. (b) Training and test loss over epochs. Test accuracy converges near 84\% after $\sim$50 epochs.}
\label{fig:crnn_training}
\end{figure}

The training and test accuracy curves for the CRNN (Fig.~\ref{fig:crnn_training}) show steady convergence, with training accuracy climbing smoothly and test accuracy plateauing around 84\% after approximately 50--60 epochs. The corresponding loss curves confirm this pattern: training loss decreases steadily while test loss stabilizes with mild fluctuation. The gap between training and test metrics indicates some degree of overfitting, which the combination of dropout and L2 regularization keeps within acceptable bounds given the dataset size.

% ============================================================
% V. DISCUSSION
% ============================================================
\section{Discussion}

\subsection{Why Sequential Beats Parallel}

The most notable architectural finding is the 8-percentage-point gap between the sequential CRNN (84\%) and the parallel CNN-RNN (76\%). Both architectures combine convolutional and recurrent components, but they do so in fundamentally different ways.

In the parallel design, the CNN and RNN branches each process the raw Mel spectrogram independently. The CNN extracts spatial/spectral features while the RNN models temporal structure, and their outputs are fused late in the pipeline. The problem is that the RNN branch must learn temporal patterns directly from 128-dimensional raw Mel frames---a task it demonstrably struggles with, as evidenced by the standalone RNN's 68\% accuracy.

In the sequential design, the CNN layers first compress and abstract the Mel spectrogram into a sequence of learned feature vectors. The LSTM then operates on this refined representation, where temporal patterns are easier to model because the convolutional preprocessing has already disentangled spectral from temporal structure. In effect, the CNN ``cleans up'' the signal before the LSTM has to make sense of it. This hierarchical composition of representations---spatial features first, temporal modeling second---is both more principled and, as our results show, substantially more effective.

\subsection{Why Deep Learning Outperforms Classical Models}

The comparison between classical and deep approaches yields a nuanced picture. The best classical models (LR and XGBoost at 71\%) match the standalone CNN's accuracy, which initially seems surprising. The explanation lies in the 51 hand-crafted features: they encode decades of domain knowledge about what makes audio signals musically meaningful (MFCCs, chroma, spectral shape). A CNN trained from scratch on raw spectrograms must rediscover these patterns, and with only 7{,}200 training samples, it cannot learn a richer representation than what domain experts have already codified.

The picture changes when temporal modeling enters. Classical features are computed as frame-level averages, discarding all information about \emph{how} features evolve over time. A song's temporal structure---its rhythmic patterns, transitions between verse and chorus, the alternating vocal phrases in Lok Dohori---is invisible to a feature vector that collapses 30 seconds into a single summary. The CRNN retains this temporal information by construction, and the 13-point accuracy gap between XGBoost (71\%) and the CRNN (84\%) is largely attributable to temporal modeling.

\subsection{Cultural Interpretation of Misclassification}

The error patterns in our CRNN model are not random; they reflect genuine cultural and acoustic affinities within Nepali music. The strongest confusion cluster links Purbeli Bhaka, Lok Dohori, and Deuda---three genres rooted in Nepali folk traditions that share instrumentation (madal drum, harmonium), vocal delivery styles, and song structures. That the model occasionally confuses these genres is less a failure than a reflection of real overlap.

Similarly, the bidirectional confusion between Pop and Aadhunik Sangeet mirrors an ongoing cultural evolution. Contemporary Nepali music increasingly adopts Western pop production techniques---synthesized instruments, four-on-the-floor drum patterns, studio-polished vocals---blurring the line between what is ``modern Nepali'' and what is ``Nepali pop.'' A strictly categorical genre taxonomy may not fully capture this continuum, and future work might benefit from soft labeling or hierarchical classification.

By contrast, Rap's very high classification accuracy (F1 = 0.93) reflects its acoustic distinctiveness even in the Nepali context: rhythmic speech delivery over sparse beats is a pattern unlike anything in the folk or modern genre categories.

\subsection{Comparison with Prior Work}

Direct comparison with published results on Western datasets is complicated by differences in genre taxonomy, dataset size, and evaluation protocol. That said, our CRNN's 84\% accuracy on an 8-genre task is broadly competitive with results reported on the 10-genre GTZAN dataset, where state-of-the-art methods typically achieve 80--90\% accuracy. This is encouraging given our relatively modest dataset size (7{,}200 training samples versus GTZAN's common augmented variants), and suggests that the CRNN architecture generalizes well to non-Western music traditions.

\subsection{Limitations}

Several limitations should be acknowledged. First, while our dataset is the first of its kind for Nepali music, it is modest by the standards of large-scale deep learning. More data would likely improve all deep models, particularly the CNN and RNN baselines. Second, the manual noise removal process, though thorough, introduces subjectivity in what constitutes ``clean'' audio. Third, the model is constrained to the eight genres it was trained on; audio from an unrepresented genre will inevitably be misclassified into one of the eight known categories. Fourth, the genre labels themselves reflect a particular taxonomic choice---other valid taxonomies exist, and some songs may legitimately belong to multiple genres. Finally, inference latency (30--40 seconds per clip on the test hardware) limits applicability in real-time scenarios.

% ============================================================
% VI. CONCLUSION
% ============================================================
\section{Conclusion and Future Work}
\label{sec:conclusion}

We have presented what is, to our knowledge, the first systematic study of automatic genre classification for Nepali music. By constructing a novel 8-genre, 8{,}000-clip dataset and evaluating nine models across two paradigms, we arrive at three principal findings.

First, the sequential CRNN architecture---which pipelines convolutional feature extraction into LSTM-based temporal modeling---achieves the best performance at 84\% accuracy, substantially outperforming both the best classical model (71\%) and the next-best deep architecture (76\%). The key to this success is hierarchical representation: the CNN learns spectral features, and the LSTM models their temporal evolution, producing a richer representation than either component can achieve alone.

Second, the choice of feature representation matters as much as the choice of model. Classical classifiers on 51 hand-crafted features match the standalone CNN on Mel spectrograms, but fall behind when temporal modeling is added. The value of deep learning for this task lies not just in learned features, but specifically in the ability to model sequential structure.

Third, misclassification patterns are culturally interpretable. Genres that share folk roots and instrumentation are more likely to be confused, while genres with globally distinctive acoustic profiles (Rap, Rock) are classified with high reliability.

Looking ahead, several directions are promising: (i) expanding the dataset through partnerships with streaming platforms and cultural archives, (ii) exploring attention mechanisms and transformer architectures that may capture long-range dependencies more effectively than LSTMs, (iii) applying data augmentation (time stretching, pitch shifting, noise injection) to improve robustness, (iv) investigating multi-label or hierarchical classification to better handle genre overlap, and (v) deploying the model as a publicly accessible tool to gather community feedback and support iterative improvement.

% ============================================================
% ACKNOWLEDGMENT
% ============================================================
\section*{Acknowledgment}

The authors thank the Department of Electronics and Computer Engineering, Purwanchal Campus, Institute of Engineering, Tribhuvan University for supporting this work. We are grateful to our project supervisor for his continuous guidance, and to the museum and radio station archivists who generously provided access to traditional music recordings.

% ============================================================
% REFERENCES
% ============================================================

\end{document}